 \documentclass[final,5p,times,twocolumn]{elsarticle}
\journal{Computer Physics Communications}
\usepackage[utf8]{inputenc}
\usepackage{braket}
\usepackage{amsmath}

\usepackage{amsbsy}
\usepackage{amstext}
\usepackage{graphicx}

\def\b#1{\mathbf{#1}}

\def\nn{\nonumber \\}

\def\fig#1{Fig.\ \ref{#1}}

\pdfoutput=1

\begin{document}

\title{Magnetotransport in a time-modulated double quantum point contact system}

\author[nuu]{Chi-Shung Tang\corref{cor1}}
\ead{cstang@nuu.edu.tw}

\author[UoI]{Kristinn Torfason}

\author[UoI]{Vidar Gudmundsson}

\cortext[cor1]{Corresponding author}
\address[nuu]{Department of Mechanical Engineering, National United
University, Miaoli 36003, Taiwan}
\address[UoI]{Science Institute, University of Iceland, Dunhaga 3,
        IS-107 Reykjavik, Iceland}

\begin{abstract}
We report on a time-dependent Lippmann-Schwinger scattering theory
that allows us to study the transport spectroscopy in a
time-modulated double quantum point contact system in the presence
of a perpendicular magnetic field. Magnetotransport properties
involving inter-subband and inter-sideband transitions are tunable
by adjusting the time-modulated split-gates and the applied magnetic
field.  The observed magnetic field induced Fano resonance feature
may be useful for the application of quantum switching.
\end{abstract}

\begin{keyword}
magnetotransport \sep time-modulated \sep quantum point contact \sep
magnetic field
\end{keyword}
\maketitle

\section{Introduction}

Electron transport in mesoscale devices smaller than the electron
phase coherence length has received extensive
studies~\cite{Beenakker1991}.  Magnetotransport and time-dependent
transport in  gate-controlled semiconducting systems are essential
fundamental entities in mesoscopic physics. Recently, It was
reported that the conductance involving Aharonov-Bohm (AB)
interference as a function of magnetic field exhibits step-like
structures~\cite{Camino2005}. Sigrit et al. measured the
differential conductance of an AB interferometer by varying the bias
voltage~\cite{Sigrist2007}. Their results indicate that varying
either the magnetic field or the electrostatic confining potentials
allows the interference to be tuned.  In this work, we investigate
the magneto-conductance in a double quantum point contact (DQPC)
system by controlling two pairs of split-gate (SG) voltages for the
manipulation of the dynamical electronic transport properties in the
DQPC-confined cavity region.

\section{Model}

\begin{figure}[htbq]
\includegraphics[width=0.35\textwidth,angle=0]{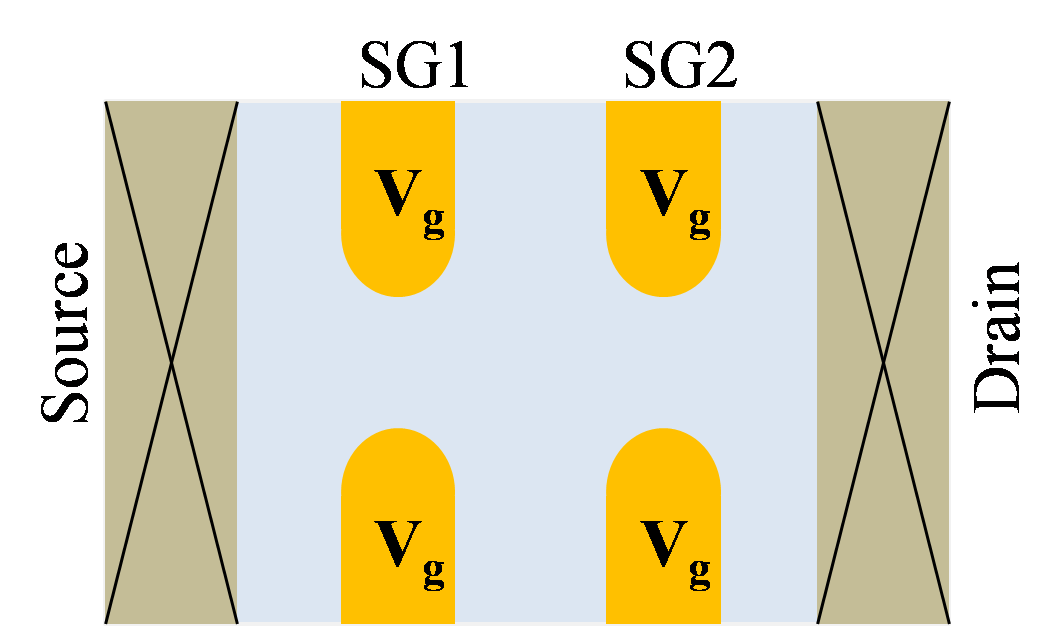}
\caption{ Schematic illustration of the double quantum point contact
system constructed by two pairs of split-gates with gate voltages
$V_g$ described by $V_{\rm sc}$ in our model. }
 \label{fig:DQPC}
\end{figure}

The system under investigation is supposed to be a parabolically
confined quantum channel fabricated from a modulation-doped
GaAs-based heterostructure with two-pairs of spit-gates defining the
DQPC system treated as a scattering potential $V_{\rm sc}(x,y,t)$,
as depicted in \fig{fig:DQPC}.
The Hamiltonian describing the system can be expressed in the form
\begin{equation}\label{eq:hamiltonian}
 {\cal H}(t) =  -\frac{\hbar^2}{2m^*} \left(\nabla^2
 - \frac{2i}{l^2} y \partial_x - \frac{y^2}{l^4} \right)
 + \frac{m^*}{2} \Omega_0^2 y^2  + V_{\rm sc}\, ,
\end{equation}
in which the effective mass $m^*$=$0.067m_e$, and the magnetic
length $l$=$\hbar/(e B)$ is related to the perpendicular magnetic
field ${\bf B}$=$B{\hat{\bf z}}$.   The characteristic confining
energy $\hbar\Omega_0$ of the parabolic confinement is modified by
the applied magnetic field leading to the effective confining energy
$\hbar\Omega_\omega = \hbar\left(\omega_c^2 +
\Omega_0^2\right)^{1/2}$ where $\omega_{\rm c}=eB/(m^*c)$.  The
scattering potential
\begin{equation}
V_{\rm sc}(x,y,t) = V_{s}(x,y) + \sum_{i=1}^2 V_{t}(x,y)\cos(\omega
t + \phi_{i})
\end{equation}
contains a static part $V_s$ as well as a time-dependent part with
strength $V_t$ and driving frequency $\omega$.  The time-modulated
SGs may have an arbitrary phase $\phi_{i}$.

We employ the mixed momentum-coordinate
representation~\cite{Gurvitz1995} to transform the total wave
function $\Psi(x,y,t)= \sum_n \phi_n(y,p) \psi_n(p,t)$ into the wave
function $\Psi(p,y,t)$ in terms of the eigenfunctions $\phi_n(y,p)$
of the unperturbed quantum channel.  Performing the expansion allows
us to obtain a coupled nonlocal time-dependent integral equation in
the momentum space:
\begin{eqnarray}\label{psi-pt}
 i\hbar\partial_t \psi_n(p,t) &=& \left[  E_n(0) + K(p)\right] \psi_n(p,t) \nonumber\\
 &&+ \sum_{n'} \int \frac{q}{2\pi} V_{n,n'}(p,q,t)\psi_{n'}(q,t)\, .
\end{eqnarray}
This equation describes the electron propagation of an asymptotic
state occupying subband $n$ along the $\b{x}$-direction from the
source electrode. Here the subband threshold $E_n(0) = \left(n +
1/2\right)\hbar\Omega_\omega$ is determined by the lateral
confinement and the effective kinetic energy $K(p) = \hbar^2
p^2(\hbar\Omega_0)^2/ [ 2m^*(\hbar\Omega_\omega)^2]$. The matrix
elements of the scattering potential
\begin{equation}\label{eq:matrix-elements}
   V_{n,n'}(p,q,t) = \int dy \ dx e^{-i(p-q)x} \phi_n^*(y,p)V(x,y,t)\phi_{n'}(y,q)
\end{equation}
indicates the electrons in the subband $n$ may be making
inter-subband transitions to the intermediate subband $n'$.

To proceed, it is convenient to transform the time-dependent wave
function from the time domain to the frequency domain
\begin{equation}
 \psi_n(p,t) = \sum_{m = -\infty}^{\infty} e^{-iE_m t/\hbar}
 \psi_n^m(p)\, ,
\end{equation}
where the quasi-energy $E_m = E_0 +m\hbar\omega$. Similarly, we have
$V_{n,n'}(p,q,t) = \sum_{m' = -\infty}^{\infty} e^{-im'\omega t}
V_{n n'}^{m'}(p,q)$ with $m'$ indicating the photon sideband index.
Defining the wave number of an electron occupying the subband $n$
and the sideband $m$ intermediate state
\begin{equation}
\frac{1}{2}\left( \frac{k_n^m}{\beta} \right)^2
\frac{(\hbar\Omega_0)^2}{\hbar\Omega_\omega} = E_m  - E_n(0)
\end{equation}
and
\begin{equation}
    \widetilde{V}_{n,n'}^{m-m'} (q,p) \equiv 2 \frac{(\hbar \Omega_{\omega})^2}{(\hbar \Omega_0)^2}
                                           \frac{\beta}{\hbar \Omega_{\omega}} V_{n,n'}^{m-m'} (q,p)
\end{equation}
 allows us to  obtain the multiple scattering identity
\begin{equation}\label{eq:def-green}
 \psi_n^m(q)=\left[ \left(\frac{k_n^m}{\beta} \right)^2 -
\left(\frac{q}{\beta} \right)^2
 \right]^{-1} \sum_{m'n'} \int \frac{dp}{2\pi} \widetilde{V}_{n,n'}^{m-m'}(q,p)\psi_{n'}^{m'}(p).
\end{equation}
Taking all the intermediate states $(n',m')$ into account, we can
obtain the Lippmann-Schwinger equation in the momentum space
\begin{eqnarray} \label{eq:lipp_schw_wavefunc}
\psi_n^m (q) &=& \psi_n^{m,0} (q) + G_n^m (q) \nonumber \\
    &&\times \frac{1}{2\pi}\sum_{n',m'} \int d\left(\frac{p}{\beta}\right)\,
            \widetilde{V}_{n,n'}^{m-m'} (q,p) \psi_{n'}^{m'}(p)
  \end{eqnarray}
in terms of the unperturbed Green function $G_n^m (q)$.  Since the
incident wave $\psi_n^{m,0}(q)$ is of the delta-function type, to
achieve exact numerical computation one has to define the $T$ matrix
 \begin{eqnarray}\label{eq:T-matrix_full}
 T_{n',n}^{m',m}(q,p) &=& V_{n',n}^{m'-m}(q,p) \\
                      &+& \sum_{r,s} \int \frac{dk}{2\pi} V_{n',r}^{m' - s}(q,k) G_r^s(k)
                      T_{r,n}^{s,m}(k,p) \nonumber
 \end{eqnarray}
that couples all the intermediate states $(r,s)$.  The potential is
expanded in the Fourier series yields a connection between the
sidebands for constructing the $T$ matrix
\begin{eqnarray}\label{eq:T-matrix_eq}
    &&T_{n',n}^{m',m} (q,p) =  V_{s,n' n}(q,p) \delta_{m'-m,0} \nn
    &&+ \frac{1}{2} V_{t, n' n} (q,p) (\delta_{m'-m,-1} + \delta_{m'-m,1}) \nn
    &&+      \sum_r \int \frac{dk}{2\pi} V_{s,n' r} (q,k) G_r^{m'} (k) T_{r,n}^{m',m} (k,p) \nn
    &&+      \frac{1}{2} \sum_r \int \frac{dk}{2\pi} V_{t, n' r}^{+} (q,k) G_r^{m'+1} (k) T_{r,n}^{(m'+1),m} (k,p) \nn
    &&+      \frac{1}{2} \sum_r \int \frac{dk}{2\pi} V_{t, n' r}^{-} (q,k) G_r^{m'-1} (k) T_{r,n}^{(m'-1),m} (k,p)
\end{eqnarray}
where $V_{t,n'r}^{\pm}(q,k) = \sum_i V_{t,n'r}(q,k) e^{\pm i
\phi_i}$ coupling the adjacent sidebands. In terms of the $T$
matrix, we can obtain the momentum-space wave function
  \begin{eqnarray}\label{eq:wave-func-t}
    \psi_{n'}^{m'}(q) &=& \psi_{n'}^{m',0}(q) + G_{n'}^{m'}(q)\nonumber\\
    &&\times \sum_{n,m} \int \frac{dk}{2\pi} T_{n',n}^{m',m}(q,k)\psi_n^{m,0}(k)\, .
  \end{eqnarray}
Performing the inverse Fourier transform to the real space and the
residue integration allows us to obtain the transmission amplitude
of the electron wave along the $\b{x}$-direction
\begin{equation}
 \b{t}_{n',n}^{m',0} = \delta_{n',n}\delta_{m',0}
 -\frac{i}{2 k_{n'}^{m'}}T_{n',n}^{m',0}\left(k_{n'}^{m'}, k_n^0\right)\, .
\end{equation}
The time-average conductance can be obtained based on the
Landauer-B\"{u}ttiker framework~\cite{Landauer1957,Buttiker1982}
\begin{equation}
G = G_0 \sum_{m'=-\infty}^{\infty}
   {\rm Tr}\left[\, \b{t}_{n',n}^{m',0}\, \left(\b{t}_{n',n}^{m',
   0}\right)^*\right]
\end{equation}
with $G_0=2e^2/h$. This indicates that the transmission matrix
connecting the contribution from all the photon sideband $m'$ of
propagating modes has to be taken into account for the electrons
occupying arbitrary subbands below the Fermi energy.

\section{Numerical Results}

We assume that the system is fabricated in a high-mobility
GaAs-based heterostructure such that the effective Rydberg energy
$E_{\mathrm{Ryd}}\approx5.9$~meV and the Bohr radius $a_{\rm
B}\approx9.8$~nm. The confining parameter of the quantum channel is
$\hbar\Omega_0=1$~meV, the length is scaled by $\beta_0^{-1}$
$\approx 33.7$~nm, and the energy is either in $\mathrm{meV}$ or in
units of $\hbar\Omega_\omega$.  The $\hbar\Omega_\omega$ =
$1.0148$~meV for the magnetic field $B$ = $0.1$~T.

The time-modulated DQPC system is described by the scattering
potential
\begin{equation}
 V_{\rm sc}(\b{r},t) = V_{\rm SG1}(\b{r},t) + V_{\rm SG2}(\b{r},t)\, ,
\end{equation}
where
\begin{equation}
 V_{\rm SG1} = V_1(t) \left[ e^{-\alpha_x (x+x_0)^2 } + e^{-\alpha_x
 (x+x_0)^2} \right] e^{- \alpha_y (y+y_0)^2}
\end{equation}
and
\begin{equation}
 V_{\rm SG2} = V_2(t)\left[ e^{-\alpha_x (x-x_0)^2} + e^{-\alpha_x
 (x-x_0)^2} \right] e^{- \alpha_y (y+y_0)^2}
\end{equation}
with $V_i(t)$ = $V_s + V_t \cos(\omega t + \phi_i)$ and $i$=$1,2$.
Moreover, we select $(\alpha_{x},\alpha_{y})$ = $(0.5,0.3)
\beta_0^2$, and $(x_0,y_0)$ = $(8,3)\beta_0$ such that the
gate-width $\sim 80$~nm and the SG-confined cavity area $\sim
540\times 200$~nm$^{2}$.

\begin{figure}[htbq]
 \includegraphics[width=0.45\textwidth]{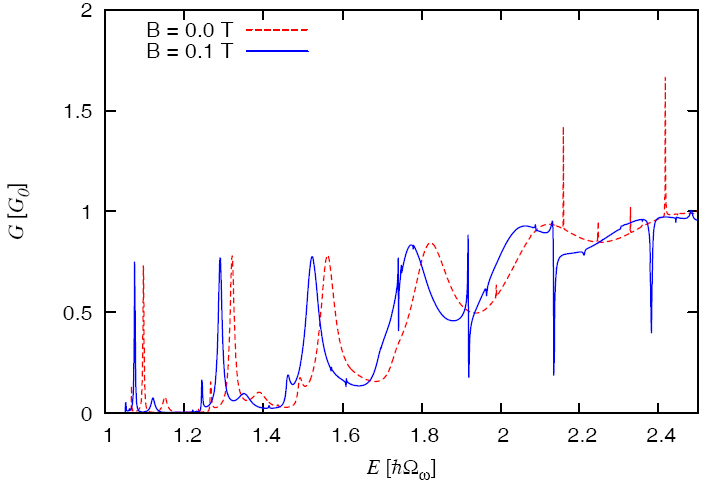}
 \caption{Conductance as a function of incident energy for the cases
 of $B = 0.0$~T (red dashed) and $B = 0.1$~T (blue solid).  The
 other parameters are $V_s = 6.0$~meV, $V_t = 1.5$~meV, $\phi = \pi$, and
 $\omega = 0.17\Omega_\omega$.}
 \label{fig2}
\end{figure}

In \fig{fig2}, we show the conductance as a function of incident
energy for the time-modulated DQPC with applied magnetic field
$B=0.1$~T (blue solid) in comparison with the zero magnetic field
situation (red dashed).  The DQPC system is confined by $V_s$ =
$6.0$~meV, and the time-modulation with strength $V_t = 1.5$~meV and
frequency $\omega = 0.17\Omega_\omega$.  In addition, we have
assumed that the phase difference between the two split-gates SG1
and SG2 is $\phi = \phi_1 - \phi_2 = \pi$. In general, the electron
kinetic energy turns out to play a role of suppressing the
quasibound state feature, namely suppressing the side-peak
structures beneath a main resonance peak in conductance. However, in
the high kinetic energy regime, an appropriate magnetic field may
induce the time-modulated Fano antiresonance features at energies
$E/\hbar\Omega_\omega \approx$ 1.75, 1.92, 2.16 as well as the
time-modulated Breit-Winger dip feature at $E/\hbar\Omega_\omega
\approx$ 2.38. Below, we focus on the robust Fano line-shape
feature: The Fano-peak is at $E/\hbar\Omega_\omega = 1.918$ and the
Fano-dip is at $E/\hbar\Omega_\omega = 1.920$, as depicted in
\fig{fig2}.

\begin{figure}[tb]
      \includegraphics[width=0.23\textwidth]{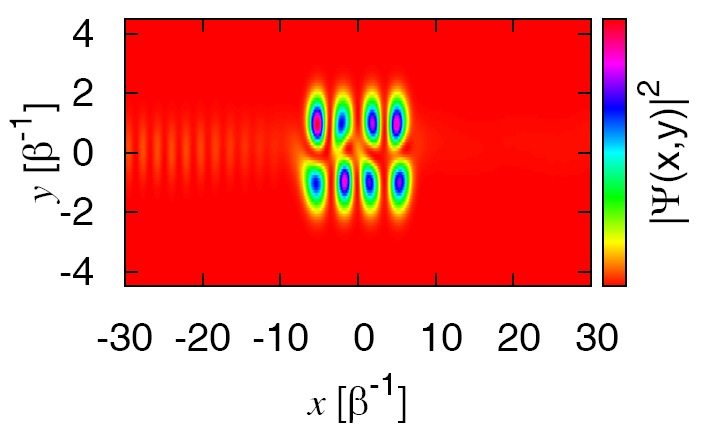}
      \includegraphics[width=0.23\textwidth]{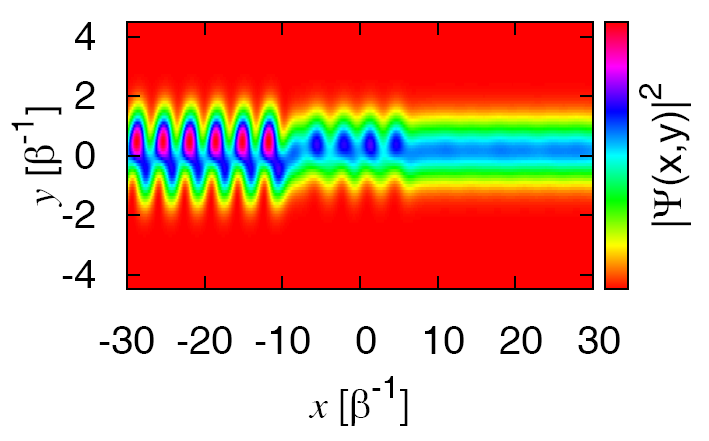}
      \caption{Probability density with magnetic field $B = 0.1$~T at the Fano peak with
      electronic energy $E/\hbar\Omega_\omega = 1.918$
      for the electron incident from the subband $n = 0$ (left) and $n = 1$ (right).}
      \label{fig3}
\end{figure}

In order to get better understanding of the magnetic-field induced
time-nodulated Fano antiresonance feature, we explore the electronic
probability density with energies around the Fano line-shape. It is
clearly shown in \fig{fig3} that the electron occupying the first
subband with higher kinetic energy favors to form a long-lived (4,2)
localized bound state in the cavity formed by the DQPC system.
However, the electron occupying the second subband with lower
kinetic energy is not fitting to the characteristic energies in the
cavity and hence forming a short-lived (4,1) extended state.  The
interference of the $n=0$ localized state and the $n=1$ extended
state induces the Fano peak at the energy $E/\hbar\Omega_\omega =
1.918$.

\begin{figure}[tb]
      \includegraphics[width=0.23\textwidth]{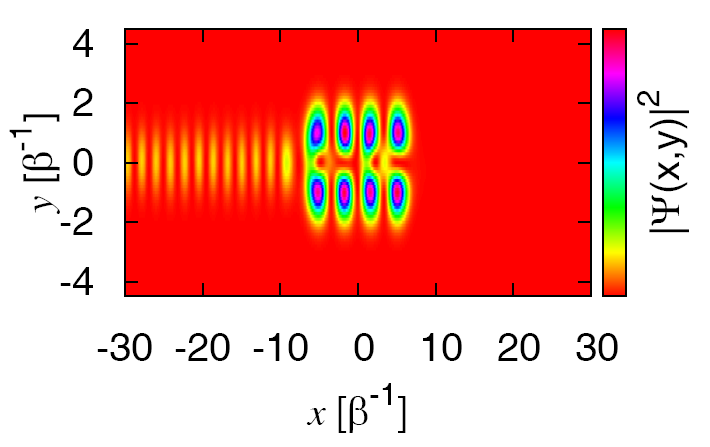}
      \includegraphics[width=0.23\textwidth]{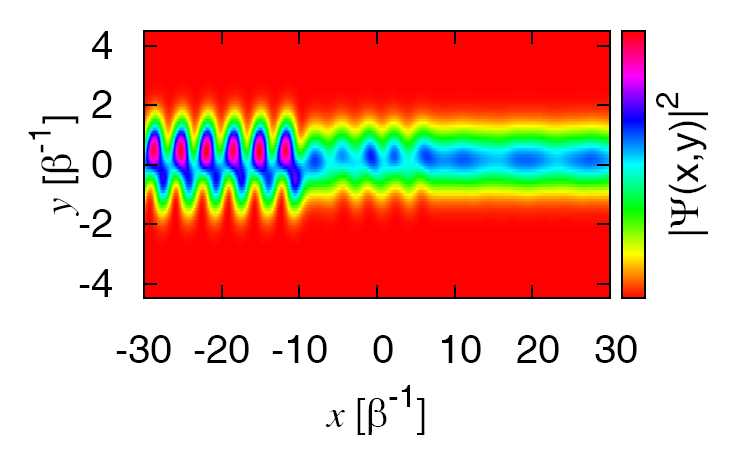}
      \caption{Probability density with magnetic field $B = 0.1$~T at the Fano dip with
      electronic energy $E/\hbar\Omega_\omega = 1.920$
      for the electron incident from the subband $n = 0$ (left) and $n = 1$ (right).}
      \label{fig4}
\end{figure}

The probability density features for the electron with incident
energy at $E/\hbar\Omega_\omega = 1.920$ are demonstrated in
\fig{fig4}.  The electrons occupying the lowest subband ($n=0$) can
also form a long-lived (4,2) localized state, but with higher
coupling to source-lead thus forming the Fano-dip line-shape.  For
the electron occupying the second subband ($n=1$) at
$E/\hbar\Omega_\omega = 1.920$, the extended (4,1) state is weaker
than the electron with $E/\hbar\Omega_\omega = 1.918$.
The energy difference $\delta E_{\rm Fano}\approx 2.03$~$\mu{\rm
eV}$ between the Fano-peak and the Fano-dip should be within the
observable resolution via the current transport measurement
technique.

\section{Summary}

In summary, we have presented coherent magnetotransport numerical
calculation on a time-modulated double QPC system and demonstrated
dynamical control of the magnetic-field induced Fano interference by
manipulating the applied magnetic field. It was reported that the
anti-symmetric ac split-gate voltage can be utilized to induce the
Fano resonance~\cite{Yang2004}.  Differently, we have reported here
by tuning an appropriate magnetic field in the DQPC system with
symmetric ac split-gates to induce the Fano resonance that becomes
non-resonant by switching off the applied magnetic field. This
robust magnetic field induced dynamic Fano resonance feature may be
useful for the magneto-control of quantum switching in arbitrary
time-modulated mesoscopic systems.

\section*{Acknowledgments}
This work was supported by the Research and Instruments Funds of the
Icelandic State; the Research Fund of the University of Iceland; the
Icelandic Science and Technology Research Programme for Postgenomic
Biomedicine, Nanoscience and Nanotechnology; and the National
Science Council of the Republic of China through Contract No.
NSC97-2112-M-239-003-MY3.



\end{document}